\documentclass[preprint,
superscriptaddress,
amsmath,amssymb,aps,showkeys,showpacs,
twoside,final,secnumarabic,
nofootinbib]{revtex4-2}

\usepackage[paperwidth=205mm,paperheight=290mm,top=17mm,bottom=25mm,
inner=17mm,outer=17mm,
twoside]{geometry}

\usepackage{cmap} 
\usepackage[T1,T2A]{fontenc}
\usepackage[utf8]{inputenc}
\usepackage[russian,english]{babel}
\usepackage{color}
\usepackage{graphicx}
\usepackage{dcolumn}
\usepackage{bm} 
\usepackage[unicode=true,colorlinks=true,linkcolor=magenta, urlcolor=blue, citecolor = blue,breaklinks]{hyperref}
\usepackage{multirow}
\usepackage{url}
\usepackage{breakurl}
\DeclareGraphicsExtensions{.eps}

\newcount\issue
\newcount\Vol
\newcount\numb
\usepackage{fancyhdr} 
\def\Vol{\textbf{80}}
\def\numb{x}
\newcommand{\z}{&&\hspace*{-1cm}}

\newcommand{\bea}{\begin{eqnarray}}
\newcommand{\eea}{\end{eqnarray}}
\newcommand{\be}{\begin{equation}}
\newcommand{\ee}{\end{equation}}

\newcommand{\ar}{a_s}

\begin{document}

\title{ CONFERENCE SECTION \\[20pt]
 QCD with Analytic Coupling: Recent Results}

\def\addresse{Veksler and Baldin Laboratory of High Energy Physics, Joint Institute for Nuclear Research, 141980 Dubna, Russia}
\def\addressb{Faculty of Physics, Moscow State University, 119991 Moscow, Russia}
\def\addressa{Bogoliubov Laboratory of Theoretical Physics, Joint Institute for Nuclear Research, 141980 Dubna, Russia}
\def\addressc{Engineering Physics Institute, Dubna State University, 141980 Dubna, Russia}
\def\addressd{Department of Physics, Universidad Tecnica Federico Santa Maria, Avenida Espana, Valparaiso 1680, Chile}

\author{\firstname{I.R.}~\surname{Gabdrakhmanov}}
\affiliation{\addresse}
\author{\firstname{N.A.}~\surname{Gramotkov}}
\affiliation{\addressb}
 \author{\firstname{A.V.}~\surname{Kotikov}}
\affiliation{\addressa}
\email[E-mail: ]{kotikov@theor.jinr..ru }
 \author{\firstname{O.V.}~\surname{Teryaev}}
\affiliation{\addressa}
 \author{\firstname{D.A.}~\surname{Volkova}}
\affiliation{\addressc}
\author{\firstname{I.A.}~\surname{Zemlyakov}}
\affiliation{\addressd}

\received{xx.xx.2025}
\revised{xx.xx.2025}
\accepted{xx.xx.2025}

\begin{abstract}
  This work provides an overview of recent results obtained in the framework of QCD with analytic coupling. Applications to the polarized Bjorken and Gross-Llewellyn Smith sum rules are considered.
\end{abstract}

\pacs{Suggested PACS}\par
\keywords{Suggested keywords   \\[5pt]}

\maketitle
\thispagestyle{fancy}


\section{Introduction}\label{intro}

According to the general principles of (local) quantum field theory (QFT) \cite{Bogolyubov:1959bfo},
observables in the spacelike region
can only have singularities for negative values of their argument $Q^2$.
However, for large $Q^2$ values, these observables are usually represented as power expansions in the running coupling constant $\alpha_s(Q^2)$,
which possesses an unphysical singularity known as the Landau pole at $Q^2 = \Lambda^2$. Therefore, to restore the analyticity of these expansions,
this pole in the strong coupling must be removed.

The strong coupling $\alpha_s(Q^2)$ satisfies the renormalization group equation
\be
L\equiv \ln\frac{Q^2}{\Lambda^2} = \int^{\overline{a}_s(Q^2)} \, \frac{da}{\beta(a)},~~ \overline{a}_s(Q^2)=\frac{\alpha_s(Q^2)}{4\pi}\,,
\label{RenGro}
\ee
with some boundary condition and the QCD $\beta$-function:
\be
\beta(\ar) ~=~ -\sum_{i=0} \beta_i \overline{a}_s^{i+2}
=-\beta_0  \overline{a}_s^{2} \, \Bigl(1+\sum_{i=1} b_i \ar^i \Bigr),~~ b_i=\frac{\beta_i}{\beta_0^{i+1}}\,, ~~
\ar(Q^2)=
\beta_0\,\overline{a}_s(Q^2)\,,
\label{beta}
\ee
where
\be
\beta_0=11-\frac{2f}{3},~~\beta_1=102-\frac{38f}{3},~~\beta_2=\frac{2857}{2}-\frac{5033f}{18}+\frac{325f^2}{54},~~
\label{beta_i}
\ee
for $f$ active quark flavors. Currently, the first five coefficients, i.e., $\beta_i$ with $i\leq 4$, are known exactly
\cite{Baikov:2016tgj,Herzog:2017ohr,Luthe:2017ttg}.
In the present work, we only require $0 \leq i\leq 2$.

Note that in Eq. (\ref{beta})
we have included the first coefficient of the QCD $\beta$-function, $\beta_0$, in the definition of $\ar$, as is customary in the analytic version of QCD
(see, e.g., Refs. \cite{ShS,MSS,BMS1,Bakulev:2006ex,Bakulev:2010gm}).

Thus, at leading order (LO), next-to-leading order (NLO), and next-to-next-to-leading order (NNLO), where $\ar(Q^2)\equiv \ar^{(1)}(Q^2)$,
$\ar(Q^2)\equiv \ar^{(2)}(Q^2)=\ar^{(1)}(Q^2)+\delta^{(2)}_s(Q^2)$ and $\ar(Q^2)\equiv \ar^{(3)}(Q^2)=\ar^{(1)}(Q^2)+\delta^{(2)}_s(Q^2)+\delta^{(3)}_s(Q^2)$, respectively,
we obtain from Eq. (\ref{RenGro})
\be
\ar^{(1)}(Q^2) = \frac{1}{L}\, ,
\delta^{(2)}_{s}(Q^2) = - \frac{b_1\ln L}{L^2} ,~~
\delta^{(3)}_{s}(Q^2) =  \frac{1}{L^3} \, \Bigl[b_1^2(\ln^2 L-\ln L-1)+b_2\Bigr]\,,
\label{asLO}
\ee
i.e., $\ar^{(i)}(Q^2)$ $(i=1,2,3)$ contain poles and other singularities at $Q^2=\Lambda^2$.

In the timelike region ($q^2 >0$),
the definition of the running coupling is rather different and its consideration is beyond the scope of this work. It can be found in the recent review \cite{Gabdrakhmanov:2025afi}.

Note that sometimes the effective argument of the strong coupling enters a region where perturbation theory (PT) becomes unreliable.
To extend the applicability of PT, some infrared modifications of the strong coupling are typically employed.
The most popular modifications are the "freezing" procedure (see, for example, Ref. \cite{Badelek:1996ap}) and the Shirkov-Solovtsov approach \cite{ShS}.

The "freezing" of the strong coupling can be implemented in a hard or soft manner.
In the hard case (see \cite{Nikolaev:1990ja}, for example), the strong coupling itself is modified: it is taken to be constant for all values of $Q^2$
less than some $Q^2_0$, i.e., $\alpha_s(Q^2)=\alpha_s(Q_0^2)$ if $Q^2 \leq Q_0^2$.

In the soft case \cite{Badelek:1996ap}, the argument of the strong coupling is modified. It involves a shift $Q^2\to Q^2 + \overline{M}^2$, where $\overline{M}$
is an additional scale (sometimes called the gluon effective mass) that significantly alters the infrared properties of $\alpha_s(Q^2)$. For massless produced quarks, the value of $\overline{M}$ is
usually taken to be the mass of the $\rho$ meson $m_{\rho}$, i.e., $ \overline{M} = m_{\rho}$.
In the case of massive quarks with mass $m_i$, the value $ \overline{M}_i = 2m_i$ is typically used.
In more complex cases, effective masses with more complicated forms are used (see, e.g., the review \cite{Deur:2016tte} with examples of masses obtained by
solving the Schwinger-Dyson equation and the papers \cite{Becher:2010tm,Becher:2012yn}, where the argument of the coupling depends on the process under consideration). Moreover, sometimes the elimination of the
Landau pole leads to additional power-law corrections (see \cite{Becher:2006mr}).

Hereafter, we will study the analytic coupling.
In Refs. \cite{ShS,MSS}, an efficient approach was developed to eliminate the Landau singularity without introducing extraneous infrared regulators,
such as the gluon effective mass (see, e.g., \cite{Badelek:1996ap,GayDucati:1993fn}).
\footnote{Numerically, couplings with an effective mass are very close to the analytic one (see \cite{Kotikov:2004uf}).}

This method is based on a dispersion relation that relates the new analytic coupling $A_{\rm MA}(Q^2)$ to the spectral function $r_{\rm pt}(s)$
obtained in the PT framework.
At LO, this gives
    \be
A^{(1)}_{\rm MA}(Q^2)
= \frac{1}{\pi} \int_{0}^{+\infty} \,
\frac{ d s }{(s + t)} \, r^{(1)}_{\rm pt}(s),~~ r^{(1)}_{\rm pt}(s)= {\rm Im} \; a_s^{(1)}(-s - i \epsilon) \,.
\label{disp_MA_LO}
\ee
The \cite{ShS,MSS} approach follows the corresponding results \cite{Bogolyubov:1959vck} obtained in Quantum Electrodynamics.

This approach is usually called the {\it Minimal Approach} (MA) (see, e.g., \cite{Cvetic:2008bn})
or {\it Analytical Perturbation Theory} (APT) \cite{ShS,MSS}.
\footnote{An overview of other similar approaches can be found in \cite{Bakulev:2008td}, including approaches \cite{Nesterenko:2003xb} that are close to APT.}

Thus, MA QCD is a very convenient approach that combines the analytical properties of QFT quantities with the results
obtained in perturbative QCD, leading to the MA coupling $A_{\rm MA}(Q^2)$,
which is close to the usual strong coupling $a_s(Q^2)$ for large $Q^2$ values but significantly different for small $Q^2$ values,
i.e., for $Q^2 \sim \Lambda^2$.

A further development of APT is the so-called fractional APT (FAPT) \cite{BMS1,Bakulev:2006ex,Bakulev:2010gm}, which extends the construction principles
described above to PT series involving non-integer powers of the coupling.
In QFT, such series arise for quantities with non-zero anomalous dimensions.
Compact expressions for quantities within FAPT were obtained mainly at LO, but this approach has also been used in higher orders,
primarily by re-expanding the corresponding couplings in powers of the LO coupling, or by using approximations.

In this review, we present the main properties of higher-order MA couplings in the FAPT framework, obtained in Refs. \cite{Kotikov:2022sos,KoZe23}
using the so-called $1/L$-expansion. Note that for the ordinary coupling, this expansion is only applicable for large $Q^2$ values, i.e., $Q^2 \gg \Lambda^2$.
However, as shown in \cite{Kotikov:2022sos,KoZe23}, the situation is quite different for analytic couplings: the $1/L$-expansion is applicable for all values of the argument. This is because the non-leading expansion corrections vanish not only
as $Q^2 \to \infty$, but also as $Q^2 \to 0$,
\footnote{The absence of high-order corrections for $Q^2 \to 0$ was also discussed in Refs. \cite{ShS,MSS}.}
resulting in only small, non-zero corrections in the region $Q^2 \sim \Lambda^2$.

Below, we present representations for the MA couplings and their (fractional) derivatives obtained in \cite{Kotikov:2022sos,KoZe23} (see also \cite{Kotikov:2022vnx}),
which are valid in principle to any PT order. However, to avoid cumbersome expressions while still demonstrating the main features of the approach from \cite{Kotikov:2022sos,KoZe23}, we restrict ourselves to the first three PT orders.

Moreover, here we show FAPT applications for describing the (polarized) Bjorken \cite{Bjorken:1966jh} and Gross-Llewellyn Smith (GLS) \cite{Gross:1969jf} sum rules.
The results presented here have been recently obtained in Refs. \cite{Gabdrakhmanov:2023rjt,Gabdrakhmanov:2024bje} and Ref. \cite{Gabdrakhmanov:GLS}, respectively.
In contrast to the analytical formulas, the results for the polarized Bjorken and GLS sum rules will be shown up to the fifth and fourth PT orders, respectively, as obtained in \cite{Kotikov:2022sos,Gabdrakhmanov:2023rjt,Gabdrakhmanov:2024bje,Gabdrakhmanov:GLS}.

The paper is organized as follows. Section 2 reviews the basic properties of the usual strong coupling and its $1/L$-expansion. Section 3 covers fractional derivatives (i.e., $\nu$-derivatives) of the usual strong coupling, whose $1/L$-expansions can be represented as operators acting on the $\nu$-derivatives of the LO strong coupling. Sections 4 and 5 present the results for the MA couplings. Sections 6 and 7 contain applications of this approach to the Bjorken and GLS sum rules, respectively. The final section provides some concluding discussions.

\section{Strong coupling}
\label{strong}

As shown in the Introduction, the strong coupling $a_s(Q^2)$ obeys the renormalization group equation (\ref{RenGro}).
For $Q^2 \gg \Lambda^2$, Eq. (\ref{RenGro}) can be solved iteratively in the form of a $1/L$-expansion
\footnote{The $1/L$-expansion provides a good approximation for the solution of Eq. (\ref{beta}) at $Q^2 \geq$ 10 GeV$^2$ (see, e.g., \cite{Shaikhatdenov:2009xd}).}
(we give the first three terms of the expansion as justified in the introduction),
which can be compactly written as
\be
a^{(1)}_{s,0}(Q^2) = \frac{1}{L_0},~~
a^{(i+1)}_{s,i}(Q^2) =
a^{(1)}_{s,i}(Q^2) + \sum_{m=2}^i \, \delta^{(m)}_{s,i}(Q^2)
\,,~~(i=0,1,2,...)\,,
\label{as}
\ee
where
\be
L_k=\ln t_k,~~t_k=\frac{1}{z_k}=\frac{Q^2}{\Lambda_k^2}\,.
\label{L}
\ee

The corrections $\delta^{(m)}_{s,k}(Q^2)$ are the same as those shown in (\ref{asLO}) with the replacement $L \to L_k$.

As shown in Eqs. (\ref{as}), in any PT order, the coupling $\ar(Q^2)$ contains the dimensional transmutation parameter $\Lambda$, which is related to the normalization of $\alpha_s(M_Z^2)$ as
\be
\Lambda_{i}=M_Z \, \exp\left\{-\frac{1}{2} \left[\frac{1}{a_s(M_Z^2)} + b_1\, \ln a_s(M_Z^2) +
\int^{\overline{a}_s(M_Z^2)}_0 \, da \, \left(\frac{1}{\beta(a)}+ \frac{1}{a^2(\beta_0+\beta_1 a)}\right)\right]\right\}\,,
\label{Lambdai}
\ee
where $\alpha_s(M_Z)=0.1178 \pm 0.0010$ from PDG24 \cite{PDG24}.

\subsection{ $f$-dependence of the coupling $\ar(Q^2)$.}

The coefficients $\beta_i$ (\ref{beta_i}) depend on the number $f$ of active quarks, which changes the coupling $\ar(Q^2)$ at thresholds $Q^2_f \sim m^2_f$, where an additional quark becomes active for $Q^2 > Q^2_f$.
Here $m_f$ is the $\overline{MS}$ mass of the $f$-quark, e.g., $m_b=4.18 +0.04-0.03$ GeV and $m_c=1.230 \pm 0.004$ GeV from PDG24 \cite{PDG24}.
\footnote{Strictly speaking, quark masses in the $\overline{MS}$ scheme depend on $Q^2$, and $m_f=m_f(Q^2=m_f^2)$. This $Q^2$-dependence is rather slow and will not be discussed here.}

Thus, the coupling $a_s$ depends on $f$, and this dependence can be incorporated into $\Lambda$, i.e., $\Lambda^f$ appears in the above Eqs. (\ref{RenGro}) and (\ref{as}).

Relations between $\Lambda_{i}^{f}$ and $\Lambda_{i}^{f-1}$, known as matching conditions between $a_s(f,Q_f^2)$ and $a_s(f-1,Q_f^2)$,
are known up to four-loop order \cite{Chetyrkin:2005ia} in the $\overline{MS}$ scheme and are usually applied at $Q_f^2=m_f^2$, where they take the simplest form (see e.g. \cite{dEnterria:2022hzv} for a recent review).

We will not consider the $f$-dependence of $\Lambda_{i}^{f}$ and $a_s(f,M_Z^2)$ here, as we mainly focus on the low-$Q^2$ region and therefore use $\Lambda_{i}^{f=3}$ $(i=0,1,2,3)$ taken from the recent Ref. \cite{Chen:2021tjz}
\footnote{The authors of \cite{Chen:2021tjz} used the PDG20 result $\alpha_s(M_Z)=0.1179(10)$. Note that very similar numerical relations for $\Lambda_i$ were also obtained by \cite{Illa} for $\alpha_s(M_Z)=0.1168(19)$ extracted by the ZEUS collaboration (see \cite{ZEUS}).}:
\be
\Lambda_0^{f=3}=142~~ \mbox{MeV},~~\Lambda_1^{f=3}=367~~ \mbox{MeV},~~\Lambda_2^{f=3}=324~~ \mbox{MeV},~~\Lambda_3^{f=3}=328~~ \mbox{MeV}\,,
\label{Lambdas}
\ee
We also use $\Lambda_4=\Lambda_3$, since in higher orders the $\Lambda_i$ values become very similar.

\begin{figure}[!htb]
\centering
\includegraphics[width=0.78\textwidth]{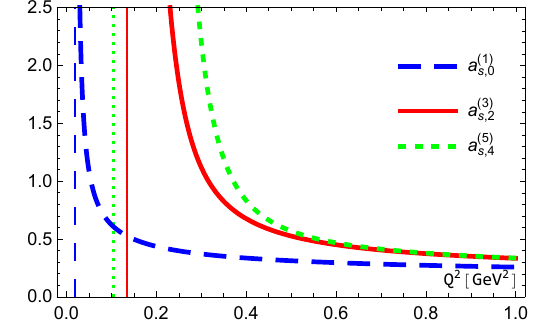}
\caption{\label{fig:as1352}
  The results for $a^{(i+1)}_{s,i}(Q^2)$ and $(\Lambda_i^{f=3})^2$ (vertical lines) with $i=0,2,4$. Here and in subsequent figures, the $\Lambda_i^{f=3}$ values from (\ref{Lambdas}) are used.
}
\end{figure}

Figure \ref{fig:as1352} shows that the strong couplings $a^{(i+1)}_{s,i}(Q^2)$ become singular at $Q^2\sim \Lambda_i^2$, and these singularities shift to higher $Q^2$ values as the PT order increases.
The values of $\Lambda_0$ and $\Lambda_j$ $(j\geq 1)$ are quite different (see Eq. (\ref{Lambdas})): the values of $(\Lambda_i^{f=3})^2$ $(i=0,2,4)$ are also indicated by vertical lines in Fig. 1.

\section{Fractional derivatives}

Following \cite{Cvetic:2006mk,Cvetic:2006gc}, we introduce the derivatives (at the $(i)$-th PT order)
\be
\tilde{a}^{(i)}_{n+1}(Q^2)=\frac{(-1)^n}{n!} \, \frac{d^n a^{(i)}_s(Q^2)}{(dL)^n} \, ,
\label{tan+1}
\ee
which are very convenient in analytical QCD (see, e.g., \cite{Kotikov:2022JETP}).

The series of derivatives $\tilde{a}_{n}(Q^2)$ can effectively replace the corresponding series of powers of $\ar$. Each derivative reduces the power of $\ar$ but introduces an additional factor of the $\beta$-function $\sim \ar^2$.
Thus, each application of a derivative effectively yields an additional factor of $\ar$, making it possible to use series of derivatives instead of series of $\ar$-powers.

At LO, the derivative series $\tilde{a}_{n}(Q^2)$ are identical to $\ar^{n}$. Beyond LO, the relationship between $\tilde{a}_{n}(Q^2)$ and $\ar^{n}$ was established in \cite{Cvetic:2006gc,Cvetic:2010di} and extended to fractional cases where $n$ is a non-integer $\nu$ in Ref. \cite{GCAK}.

Now consider the $1/L$-expansion of $\tilde{a}^{(k)}_{\nu}(Q^2)$. We can raise the results (\ref{as}) to the power $\nu$ and then reconstruct $\tilde{a}^{(k)}_{ \nu}(Q^ 2)$ using the relations between $\tilde{a}_{\nu}$ and $\ar^{\nu}$ obtained in \cite{GCAK}. This procedure is described in more detail in Appendix A of \cite{Kotikov:2022vnx}.
Here we present only the final results, which take the form
\footnote{The expansion (\ref{tdmp1N}) is similar to those used in Refs. \cite{BMS1,Bakulev:2006ex} for expanding ${\bigl({a}^{(i+1)} _{s,i}(Q^2)\bigr)}^ {\nu}$ in powers of $a^{(1)}_{s,i}(Q^2)$.}:
\bea
\z\tilde{a}^{(1)}_{\nu,0}(Q^2)={\bigl(a^{(1)}_{s,0}(Q^2)\bigr)}^{\nu} = \frac{1}{L_0^{\nu}},~
\tilde{a}^{(i+1)}_{\nu,i}(Q^2)=\tilde{a}^{(1)}_{\nu,i}(Q^2) + \sum_{m=1}^{i}\, C_m^{\nu+m}\, \tilde{\delta}^{(m+1)}_{\nu,i}(Q^2),~~\nonumber\\
\z\tilde{\delta}^{(m+1)}_{\nu,i}(Q^2)=
\hat{R}_m \, \frac{1}{L_i^{\nu+m}},~~C_m^{\nu+m}=\frac{\Gamma(\nu+m)}{m!\Gamma(\nu)}\,,
\label{tdmp1N}
\eea
where
\be
\hat{R}_1=b_1 \Bigl[\hat{Z}_1(\nu)+ \frac{d}{d\nu}\Bigr],~~
\hat{R}_2=b_2 + b_1^2 \Bigl[\frac{d^2}{(d\nu)^2} +2 \hat{Z}_1(\nu+1)\frac{d}{d\nu} + \hat{Z}_2(\nu+1 )\Bigr]
\,
\label{hR_i}
\ee
and $\hat{Z}_j(\nu)$ $(j=1,2)$ are combinations of Euler $\Psi$-functions and their derivatives (see Appendix A of \cite{Kotikov:2022vnx}).

The representation (\ref{tdmp1N}) of the $\tilde{\delta}^{(m+1)}_{\nu,i}(Q^2)$ corrections as $\hat{R} _m$-operators is very important\footnote{The results for $\hat{R}_m$-operators reflect a transcendental principle \cite{Kotikov:2000pm}: the functions $\hat{Z}_k(\nu)$ ($k \leq m$) contain Polygamma functions $\Psi_k(\nu)$ and their products, such as $\Psi_{k-l}(\nu)\Psi_l(\nu)$, and also products with more factors, all with the same total index $k$. However, the significance of this property is not yet clear.}
and allows us to similarly present higher-order results for the $1/L$-expansion of analytic couplings.

\section{MA couplings}

We first present the LO results, then proceed beyond LO following our results (\ref{tdmp1N}) and (\ref{hR_i}) for the ordinary strong coupling from the previous section.

\subsection{LO}

The LO MA coupling $A^{(1)}_{{\rm MA},\nu,0}$ has the following form \cite{BMS1}
\be
A^{(1)}_{{\rm MA},\nu,0}(Q^2) = {\left( a^{(1)}_{\nu,0}(Q^2)\right)}^{\nu} - \frac{{\rm Li}_{1-\nu}(z_0)}{\Gamma(\nu)}=
\frac{1}{L_0^{\nu}}- \frac{{\rm Li}_{1-\nu}(z_0)}{\Gamma(\nu)} \equiv \frac{1}{L_0^{\nu}}-\Delta^{(1)}_{\nu,0}\,,
\label{tAMAnu}
\ee
where
\be
   {\rm Li}_{\nu}(z)=\sum_{m=1}^{\infty} \, \frac{z^m}{m^{\nu}} 
   \label{Linu}
\ee
is the Polylogarithm.

For $\nu=1$, we recover the famous Shirkov-Solovtsov result \cite{ShS}:
\be
\hspace{-0.5cm} A^{(1)}_{\rm MA,0}(Q^2) \equiv A^{(1)}_{\rm MA,\nu=1,0}(Q^2)
=\frac{1}{L_0}- \frac{z_0}{1-z_0}\,.
\label{tAM1}
\ee
Note that the result (\ref{tAM1}) can be derived directly from the integral form (\ref{disp_MA_LO}), as was done in Ref. \cite{ShS}.

\subsection{Beyond LO}

Following Eq. (\ref{tAMAnu}) for the LO analytic coupling, we consider the derivatives of the MA coupling, defined as
\be
\tilde{A}_{{\rm MA},n+1}(Q^2)=\frac{(-1)^n}{
  n!} \, \frac{d^n A_{\rm MA}(Q^2)}{(dL)^n}\, .
\label{tanMA+1}
\ee

By analogy with the ordinary coupling, and using the results (\ref{tdmp1N}), we obtain for the MA analytic coupling $\tilde{A}^{(i+1)}_{{\rm MA},\nu,i}$ the following expression:
\be
\tilde{A}^{(i+1)}_{{\rm MA},\nu,i}(Q^2) = \tilde{A}^{(1)}_{{\rm MA},\nu,i}(Q^2) + \sum_{m=1}^{i}  \, C^{\nu+m}_m \tilde{\delta}^{(m+1)}_{{\rm A},\nu,i}(Q^2),
\label{tAiman}
\ee
where $\tilde{A}^{(1)}_{{\rm MA},\nu,i}$ is given in Eq. (\ref{tAMAnu}) and
\be
\tilde{\delta}^{(m+1)}_{{\rm A},\nu,i}(Q^2)= \tilde{\delta}^{(m+1)}_{\nu,i}(Q^2) -  \hat{R}_m \left( \frac{{\rm Li}_{-\nu-m+1}(z_i)}{\Gamma(\nu+m)}\right),~
\label{tdAman}
\ee
with $\tilde{\delta}^{(m+1)}_{\nu,i}(Q^2)$ and $\hat{R}_m$ given in Eqs. (\ref{tdmp1N}) and (\ref{hR_i}), respectively.

The relations (\ref{tAMAnu}) reflect the fact that the MA procedure (\ref{tAMAnu}) commutes with the operation $d/(d\nu)$.
Thus, to obtain (\ref{tAMAnu}), we assume that the form (\ref{tdmp1N}) for the usual coupling $a_s$ at higher orders applies exactly in the same way to the MA coupling.

After some calculations, we obtain the following expressions without explicit operators:
\be
\tilde{\Delta}^{(i+1)}_{\nu,i}
=\tilde{\Delta}^{(1)}_{\nu,i}
+\sum_{m=1}^i C_m^{\nu+m} \, \overline{R}_m(z_i) \left( \frac{{\rm Li}_{-\nu-m+1}(z_i)}{\Gamma(\nu+m)}\right)
\, ,
\label{tAMAnu.2}
\ee
where
\bea
&&\overline{R}_1(z)=b_1\Bigl[\overline{\gamma}_{\rm E}
  +{\rm M}_{-\nu,1}(z)\Bigr], \nonumber \\
&&\overline{R}_2(z)=b_2 + b_1^2\Bigl[{\rm M}_{-\nu-1,2}(z) + 2\overline{\gamma}_{\rm E}
  {\rm M}_{-\nu-1,1}(z) + \overline{\gamma}^2_{\rm E}
   - \zeta_2\Bigr]
\label{oRi}
\eea
with the Euler constant $\gamma_{\rm E}$ and
\be
 \overline{\gamma}_{\rm E}=\gamma_{\rm E}-1,~~  {\rm Li}_{\nu,k}(z)=(-1)^k\,\frac{d^k}{(d\nu)^k}  \,{\rm Li}_{\nu}(z) =
   \sum_{m=1}^{\infty} \, \frac{z^m\ln^k m}{m^{\nu}},~~{\rm M}_{\nu,k}(z)=\frac{{\rm Li}_{\nu,k}(z)}{{\rm Li}_{\nu}(z)} \, .
   \label{Mnuk}
\ee
We see that the $\Psi(\nu)$-function and its derivatives have completely canceled out.
Note that another form for $\tilde{\Delta}^{(m+1)}_{\nu,i}(Q^2)$ is given in Appendix B of \cite{Kotikov:2022sos}.

Thus, for the MA analytic couplings $\tilde{A}^{(i+1)}_{{\rm MA},\nu}$, we have the expressions:
\be
\tilde{A}^{(i+1)}_{{\rm MA},\nu,i}(Q^2) = \tilde{A}^{(1)}_{{\rm MA},\nu,i}(Q^2) + \sum_{m=1}^{i}  \, C^{\nu+m}_m \tilde{\delta}^{(m+1)}_{{\rm A},\nu,i}(Q^2) \,
\label{tAiman}
\ee
where
\bea
&&\tilde{A}^{(1)}_{{\rm A},\nu,i}(Q^2) = \tilde{a}^{(1)}_{\nu,i}(Q^2) -  \frac{{\rm Li}_{1-\nu}(z_i)}{\Gamma(\nu)},
~~\nonumber \\
&&\tilde{\delta}^{(m+1)}_{{\rm A},\nu,i}(Q^2)= \tilde{\delta}^{(m+1)}_{\nu,i}(Q^2) -  \overline{R}_m(z_i)   \, \frac{{\rm Li}_{-\nu+1-m}(z_i)}{\Gamma(\nu+m)}
\label{tdAman}
\eea
and $\tilde{\delta}^{(k+1)}_{\nu,m}(Q^2)$ are given in Eq. (\ref{tdmp1N}).

\subsection{The case $\nu=1$
}

Here we present only the results for the case $\nu=1$:
\be
A^{(i+1)}_{{\rm MA},i}(Q^2)\equiv \tilde{A}^{(i+1)}_{{\rm MA},\nu=1,i}(Q^2) = A^{(1)}_{{\rm MA},i}(Q^2) + \sum_{m=1}^{i}  \, \tilde{\delta}^{(m+1)}_{{\rm A},\nu=1,i}(Q^2),
\label{tAiman.1}
\ee
where $A^{(1)}_{{\rm MA},i}(Q^2)$
is shown in Eq. (\ref{tAM1}) and
\be
\tilde{\delta}^{(m+1)}_{{\rm A},\nu=1,i}(Q^2)
= \tilde{\delta}^{(m+1)}_{\nu=1,i}(Q^2)- \frac{P_{m,1}(z_i)}{m!} \, ,
\label{tdAmanA}
\ee
with 
\bea
\z
P_{1,\nu}(z)=b_1\Bigl[\overline{\gamma}_{\rm E}
  {\rm Li}_{-\nu}(z)+{\rm Li}_{-\nu,1}(z)\Bigr],~~
\nonumber \\
\z P_{2,\nu}(z)=b_2 \,{\rm Li}_{-\nu-1}(z) + b_1^2\Bigl[{\rm Li}_{-\nu-1,2}(z) + 2\overline{\gamma}_{\rm E}
  {\rm Li}_{-\nu-1,1}(z)
  +  \Bigl(\overline{\gamma}^2_{\rm E}-
  \zeta_2\Bigr) \, {\rm Li}_{-\nu-1}(z) \Bigr]\,,
\label{Pkz}
\eea
and
\be
 {\rm Li}_{n,m}(z)= \sum_{m=1} \, \frac{\ln^k m}{m^n},~~
    {\rm Li}_{-1}(z)= \frac{z}{(1-z)^2},~~{\rm Li}_{-2}(z)= \frac{z(1+z)}{(1-z)^3}
    \, .
\label{Lii.1}
\ee


\section{The behaviour of MA couplings}

\begin{figure}[!htb]
\centering
\includegraphics[width=0.58\textwidth]{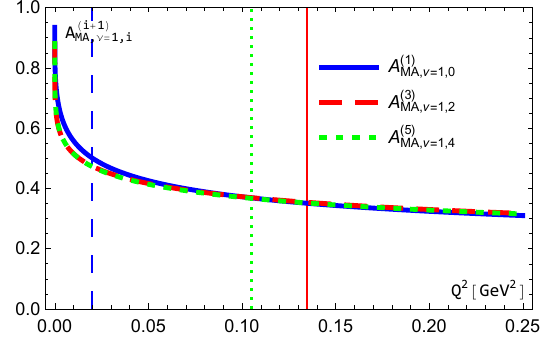}
\caption{\label{fig:A123}
  The results for $A^{(i+1)}_{\rm MA,\nu=1,i}(Q^2)$
  and $(\Lambda_i^{f=3})^2$ (vertical lines)
  with $i=0,2,4$.}
\end{figure}

\begin{figure}[!htb]
\centering
\includegraphics[width=0.58\textwidth]{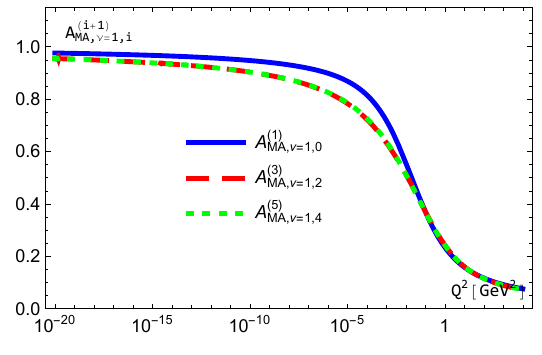}
    \caption{\label{fig:A123LOG}
      The results for $A^{(i+1)}_{\rm MA,\nu=1,i}(Q^2)$ ($i=0,1,2$)  but with the logarithmic scale.
    }
\end{figure}

\begin{figure}[!htb]
\centering
\includegraphics[width=0.58\textwidth]{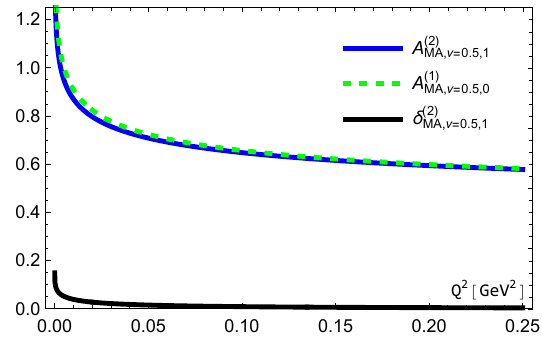}
    \caption{\label{fig:Afrac3}
      The results for $A^{(1)}_{\rm MA,\nu=0.5,0}(Q^2)$, $A^{(2)}_{\rm MA,\nu=0.5,1}(Q^2)$ and $\delta^{(2)}_{\rm A,\nu=0.5,1}(Q^2)$.
    }
\end{figure}

\begin{figure}[!htb]
\centering
\includegraphics[width=0.58\textwidth]{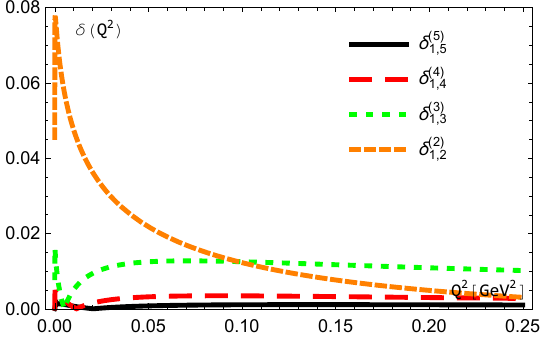}
    \caption{\label{fig:deltaALL}
      The results for $\delta^{(i+1)}_{\rm A,\nu=1,i}(Q^2)$ with $i=1,2,3,4$.}
\end{figure}

Here we study the behaviour of the MA coupling
$A^{(k+1)}_{{\rm MA},k}(Q^2)=\tilde{A}^{(k+1)}_{{\rm MA},\nu=1,k}(Q^2)$.
%
In Figs. \ref{fig:A123} and \ref{fig:A123LOG} we see that the differences between $A^{(i+1)}_{\rm MA,\nu=1,i}(Q^2)$ with $i=0,2,4$
are rather small and have nonzero values only around the
position $Q^2=\Lambda_i^2$.
In Fig. \ref{fig:A123} the values of $(\Lambda_i^{f=3})^2$ $(i=0,2,4)$ are shown by vertical lines (as it was in Fig. 1).

Fig. \ref{fig:Afrac3}
shows the results for  $A^{(1)}_{\rm MA,\nu,0}(Q^2)$ and $A^{(2)}_{\rm MA,\nu,1}(Q^2)$ and their difference
$\delta^{(2)}_{\rm A,\nu,1}(Q^2)$, which is strongly less then the couplings
themselves. 
Also Fig. \ref{fig:deltaALL} shows that the differences $\delta^{(i+1)}_{\rm A,\nu=1,i}(Q^2)$
$(i\geq 2)$ essentially less then $\delta^{(2)}_{\rm A,\nu=1,1}(Q^2)$.

Thus, 
contrary to the case of the usual coupling,
considered in Fig. 1, the $1/L$-expansion of the MA coupling
is very good approximation at any $Q^2$ values.
Moreover, the differences between $A^{(i+1)}_{\rm MA,\nu=1,i}(Q^2)$ and $A^{(1)}_{\rm MA,\nu=1,0}(Q^2)$ are very small.
Thus, the expansions of  $A^{(i+1)}_{\rm MA,\nu=1,i}(Q^2)$ $i\geq 1$ through the LO coupling
$A^{(1)}_{\rm MA,\nu=1,0}(Q^2)$ done in Refs. \cite{BMS1,Bakulev:2006ex,Bakulev:2010gm} are
very good approximations.
Also
the approximation
\be
A^{(i+1)}_{\rm MA,\nu=1,i}(Q^2)=A^{(1)}_{\rm MA,\nu=1,0}(k_iQ^2),~~(i=1,2)\,,
\label{Appro}
\ee
introduced in \cite{Pasechnik:2008th,Khandramai:2011zd} and used in \cite{Kotikov:2010bm} is very convenient, too.



Note that the results (\ref{tAiman}) for the analytic constraint $\tilde{A}^{(k+1)}_{{\rm MA},\nu,k}(Q^2)$ are very convenient
for both large and small values of $Q^2$. For $Q^2 \sim \Lambda_i^2$, both sides, the standard strong coupling and the additional term,
have singularities that sum to cancel. Therefore, some numerical applications of the results (\ref{tAiman}) may be difficult.
Note that there is an additional form that is very useful for $Q^2 \sim \Lambda_i^2$ and can also be used for any values of $Q^2$,
except for the ranges of very large and very small values of $Q^2$. Its consideration together with integral forms that extended
the LO form (\ref{disp_MA_LO}) beyond LO is not the subject of this review and can be found in \cite{Gabdrakhmanov:2025afi,Kotikov:2022sos}.

\section{Bjorken and Gross-Llewellyn Smith sum rules}

The polarized Bjorken sum rule  \cite{Bjorken:1966jh} 
is defined as the difference between the proton and neutron polarized SFs,
integrated over the entire interval $x$
\footnote{
  The integrals $\Gamma_1^{p}(Q^2)$ and $\Gamma_1^{n}(Q^2)$ themselves were studied in \cite{Pasechnik:2010fg} but such a study is
    beyond the scope of this review.}
\be
\Gamma_1^{p-n}(Q^2)=\int_0^1 \, dx\, \bigl[g_1^{p}(x,Q^2)-g_1^{n}(x,Q^2)\bigr].
\label{Gpn} 
\ee

The GLS sum rule is defined by the integral \cite{Gross:1969jf}
\bea
C_{GLS}^{p+n}(Q^2)=\frac{1}{2}\, \int\limits^1_0\left[F_3^{\overline{\nu}p}(x,Q^2)+F_3^{\nu n}(x,Q^2)\right]dx\,,
 \label{GLS.I}
\end{eqnarray}
which in the quark-parton model counts the number of valence quarks in the proton.

Perturbatively computed radiative corrections for the Bjorken and Gross-Llewellyn-Smith sum rules serve as an important test of the suitability
of QCD as a theory of strong interactions.

Theoretically, the quantities $\Gamma_1^{p-n}(Q^2)$ and $C_{GLS}^{p+n}(Q^2)$
can be written in the 
Operator Product Expansion (OPE)
forms
(see Ref. \cite{Shuryak:1981pi,Balitsky:1989jb})
\bea
&&\Gamma_1^{p-n}(Q^2)=
\frac{g_A}{6} \, \bigl(1-D_{\rm BS}(Q^2)\bigr) +\sum_{i=1}\frac{\mu^{\rm BS}_{2i+2}}{Q^{2i}} \, ,
\label{Gpn.OPE}\\
&&C_{GLS}^{p+n}(Q^2)=
3 \, \bigl(1-D_{\rm GLS}(Q^2)\bigr) +\sum_{i=1}\,\frac{\mu^{\rm GLS}_{2(i+1)}}{Q^{2n}} \, ,
\label{Gpn.OPEab}
\eea
where $g_A$=1.2762 $\pm$ 0.0005 \cite{PDG24} is
the nucleon axial charge, $(1-D_{BS}(Q^2))$ and $(1-D_{\rm GLS}(Q^2)$ are the leading-twist (or twist-two)
contributions, and $\mu^{\rm BS}_{2i+2}/Q^{2i}$ and $\mu^{\rm GLS}_{2i+2}/Q^{2i}$ $(i\geq 1)$ are the higher-twist (HT)
contributions.

Since we plan to consider in particular small $Q^2$ values here,
the HT representations (\ref{Gpn.OPE}) and (\ref{Gpn.OPEab}) are not so convenient. So,
it is preferable to use 
the so-called "massive" twist-four representations, which includes a part of the HT
contributions of (\ref{Gpn.OPE}) and (\ref{Gpn.OPEab}) (see Refs. \cite{Teryaev:2013qba,Gabdrakhmanov:2017dvg}):
\bea
\Gamma_1^{p-n}(Q^2)=
\frac{g_A}{6} \, \bigl(1-D_{\rm BS}(Q^2)\bigr) +\frac{\hat{\mu}^{\rm BS}_4 M_{\rm BS}^2}{Q^{2}+M_{\rm BS}^2} \, ,
\label{Gpn.mOPE}\\
C_{GLS}^{p+n}(Q^2)=
3 \, \bigl(1-D_{GLS}(Q^2)\bigr) +\frac{\hat{\mu}^{GLS}_4 M_{GLS}^2}{Q^{2}+M_{GLS}^2} \, .
\label{Gpn.mOPEa}
\eea
where in the Bjorken sum rule case the values of $\hat{\mu}^{\rm BS}_4$ and $M_{\rm BS}^2$ have been fitted in Refs. \cite{Ayala:2017uzx,Ayala:2018ulm}
in the different analytic QCD models.
For $Q^2 >>M_{\rm a}^2$ $(a=BS,GLS)$ , the "massive" twist-four representation
can be expanded in powers of $M^2_{\rm a}/Q^2$, and the obtained results will have the form shown on  the right-hand side of (\ref{Gpn.OPE}) and (\ref{Gpn.OPEab}).

Up to the $k$-th PT order, the twist-two
parts have the form
\be
D^{(1)}_{\rm a}(Q^2)=\frac{4}{\beta_0} \, a^{(1)}_s,~~D^{(k\geq2)}_{\rm a}(Q^2)=\frac{4}{\beta_0} \, a^{(k)}_s\left(1+\sum_{m=1}^{k-1} d^{\rm a}_m \bigl(a^{(k)}_s\bigr)^m
\right),~~({\rm a=BS, GLS}),
\label{DBS} 
\ee
where $d^{\rm a}_1$, $d^{\rm a}_2$ and $d^{\rm a}_3$ are known from exact calculations (see \cite{Baikov:2010je} and references therein).
The $d^{\rm BS}_4$ value 
was  estimated  in Ref. \cite{Ayala:2022mgz}.
\footnote{The quantities $D^{(k\geq2)}_{\rm a}(Q^2)$ also contain the contributions of heavy quarks, calculated in \cite{Blumlein:2016xcy}.
A study of corrections within the framework of analytical QCD can be found in \cite{Gabdrakhmanov:2024vol,Gabdrakhmanov:2024bje} and \cite{Gabdrakhmanov:GLS},
for the Bjorken and GLS sum rules, respectively. However, this topic is beyond the scope of this review.}

Converting the coupling
powers into its derivatives, we have
\be
D^{(1)}_{\rm a}(Q^2)=\frac{4}{\beta_0} \, \tilde{a}^{(1)}_1,~~D^{(k\geq2)}_{\rm a}(Q^2)=
\frac{4}{\beta_0} \, \left(\tilde{a}^{(k)}_{1}+\sum_{m=2}^k\tilde{d}^{\rm a}_{m-1}\tilde{a}^{(k)}_{m}\right),~~({\rm a=BS, GLS}), 
\label{DBS.1} 
\ee
where
\bea
&&\tilde{d}^{\rm a}_1=d^{\rm a}_1,~~\tilde{d}^{\rm a}_2=d^{\rm a}_2-b_1d^{\rm a}_1,~~\tilde{d^{\rm a}}_3=
d^{\rm a}_3-\frac{5}{2}b_1d^{\rm a}_2-\bigl(b_2-\frac{5}{2}b^2_1\bigr)\,d^{\rm a}_1,\nonumber \\
&&\tilde{d}^{\rm a}_4=d^{\rm a}_4-\frac{13}{3}b_1d^{\rm a}_3 -\bigl(3b_2-\frac{28}{3}b^2_1\bigr)\,d^{\rm a}_2-\bigl(b_3-\frac{22}{3}b_1b_2+\frac{28}{3}b^3_1\bigr)\,d^{\rm a}_1
\label{tdi} 
\eea
with
$b_i=\beta_i/\beta_0^{i+1}$.

In MA QCD, the results (\ref{Gpn.mOPE}) become as follows
\bea
&&\Gamma_{\rm{MA},1}^{p-n}(Q^2)=
\frac{g_A}{6} \, \bigl(1-D_{\rm{MA,BS}}(Q^2)\bigr) +\frac{\hat{\mu}^{\rm BS}_{\rm{MA},4}M_{\rm BS}^{2}}{Q^{2}+M_{\rm BS}^2},~~
\label{Gpn.MA} \\
&&C_{\rm MA,GLS}^{p+n}(Q^2)=
3 \, \bigl(1-D_{\rm{MA,GLS}}(Q^2)\bigr) +\frac{\hat{\mu}^{\rm GLS}_{\rm{MA},4}M_{\rm GLS}^{2}}{Q^{2}+M_{\rm GLS}^2},~~
\label{Gpn.MAa}
\eea
where the perturbative parts $D_{\rm{BS,MA}}(Q^2)$ and $D_{\rm{GLS,MA}}(Q^2)$
take
the same forms, however, with analytic coupling
$\tilde{A}^{(k)}_{\rm MA,\nu}$ (the corresponding expressions are taken from \cite{Kotikov:2022sos})
\be
D^{(1)}_{\rm MA,a}(Q^2)=\frac{4}{\beta_0} \, A_{\rm MA}^{(1)},~~
D^{k\geq2}_{\rm{MA,a}}(Q^2) =\frac{4}{\beta_0} \, \Bigl(A^{(1)}_{\rm MA}
+ \sum_{m=2}^{k} \, \tilde{d}^{\rm a}_{m-1} \, \tilde{A}^{(k)}_{\rm MA,\nu=m} \Bigr)\,,~~({\rm a=BS, GLS}),
\label{DBS.ma}
\ee


For the case of 3 active quark flavors ($f=3$), which is accepted in this paper (see discussions in Section 2), we have
and
\bea
&& d^{\rm BS}_1=\tilde{d}^{\rm BS}_1=1.59,~d^{\rm BS}_2=3.99~(\tilde{d}^{\rm BS}_2=2.73),~d^{\rm BS}_3=15.42~(\tilde{d}^{\rm BS}_3=8.61), \nonumber \\
&& d^{\rm BS}_4=63.76~(\tilde{d}^{\rm BS}_4=21.52),
\label{td123}\\
&&d^{\rm GLS}_1=\tilde{d}^{\rm GLS}_1=1.59,~~d^{\rm GLS}_2=3.75~(\tilde{d}^{\rm GLS}_2=2.51),~~d^{\rm GLS}_3=16.77~(\tilde{d}^{\rm GLS}_3=10.44),
\label{td123a} 
\eea
i.e., the coefficients in the series of derivatives are slightly smaller.

It is important to emphasize that the perturbative structure of the Bjorken sum rule is very similar to that of the GLS sum rule. Both sum rules are governed
by the identical (in LO) twist-2 part. Their approximate equivalence has been discussed in the literature (see \cite{Broadhurst:1993ru}).
\footnote{The similarity of perturbative parts for the structure functions $F_3(x,Q^2)$ and $g_1(x,Q^2)$ themselves has been explored in Ref. \cite{Kotikov:1996vr}.}
Therefore, building on the definitions of
the perturbative coefficients $d_i$, we can (with some precision) treat BSR and GLS as the same dependence with only different general factor:
\be
C_{GLS}^{p+n}(Q^2)\approx \frac{18}{g_A}\,\Gamma_{1}^{p-n}(Q^2)\,
\label{GLS.app}
\ee
and, thus, for analytic QCD we have similarly
\be
C_{\rm{A},GLS}^{p+n}(Q^2)\approx \frac{18}{g_A}\,\Gamma_{\rm{A},1}^{p-n}(Q^2)\,.
\label{GLS.app.ma}
\ee

\subsection{Results for Bjorken sum rule}

\begin{table}[t]
\begin{center}
\begin{tabular}{|c|c|c|c|}
\hline
& $M_{\rm BS}^2$ [GeV$^2$] for $Q^2 \leq 5$ GeV$^2$ & $\hat{\mu}^{\rm BS}_{\rm{MA},4}$  for $Q^2 \leq 5$ GeV$^2$& $\chi^2/({\rm d.o.f.})$ for $Q^2 \leq 5$ GeV$^2$ \\
& (for $Q^2 \leq 0.6$ GeV$^2$) & (for $Q^2 \leq 0.6$ GeV$^2$) & (for $Q^2 \leq 0.6$ GeV$^2$) \\
 \hline
 LO & 0.472 $\pm$ 0.035 & -0.212 $\pm$ 0.006 & 0.667  \\
 & (1.631 $\pm$ 0.301) & (-0.166 $\pm$ 0.001) & (0.789)  \\
 \hline
 NLO & 0.414 $\pm$ 0.035 & -0.206 $\pm$ 0.008 & 0.728  \\
& (1.545 $\pm$ 0.287) & (-0.155 $\pm$ 0.001) & (0.757)  \\
 \hline
 N$^2$LO & 0.397 $\pm$ 0.034 & -0.208$\pm$ 0.008 & 0.746  \\
 & (1.417 $\pm$ 0.241) & (-0.156 $\pm$ 0.002) & (0.728)  \\
 \hline
 N$^3$LO & 0.394 $\pm$ 0.034 & -0.209 $\pm$ 0.008 & 0.754  \\
  & (1.429 $\pm$ 0.248) & (-0.157 $\pm$ 0.002) & (0.747)  \\
   \hline
N$^4$LO & 0.397 $\pm$ 0.035 & -0.208 $\pm$ 0.007 & 0.753  \\
 & (1.462 $\pm$ 0.259) & (-0.157 $\pm$ 0.001) & (0.754)  \\
 \hline
\end{tabular}
\end{center}
\caption{%
  The values of the fit parameters in (\ref{Gpn.MA}).
}
\label{Tab:BSR}
\end{table}

\begin{figure}[t]
\centering
\includegraphics[width=0.98\textwidth]{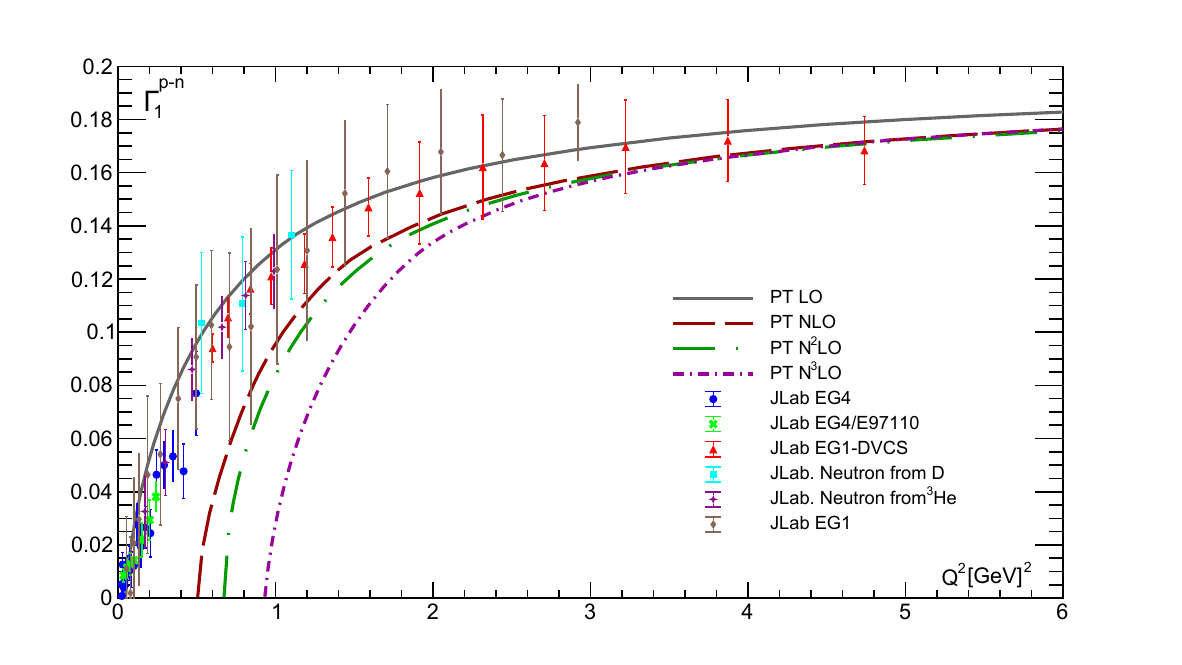}
\caption{
  \label{fig:PT}
  The results for $\Gamma_1^{p-n}(Q^2)$ in the first  four
  orders of PT.
    }
\end{figure}

\begin{figure}[t]
\centering
\includegraphics[width=0.98\textwidth]{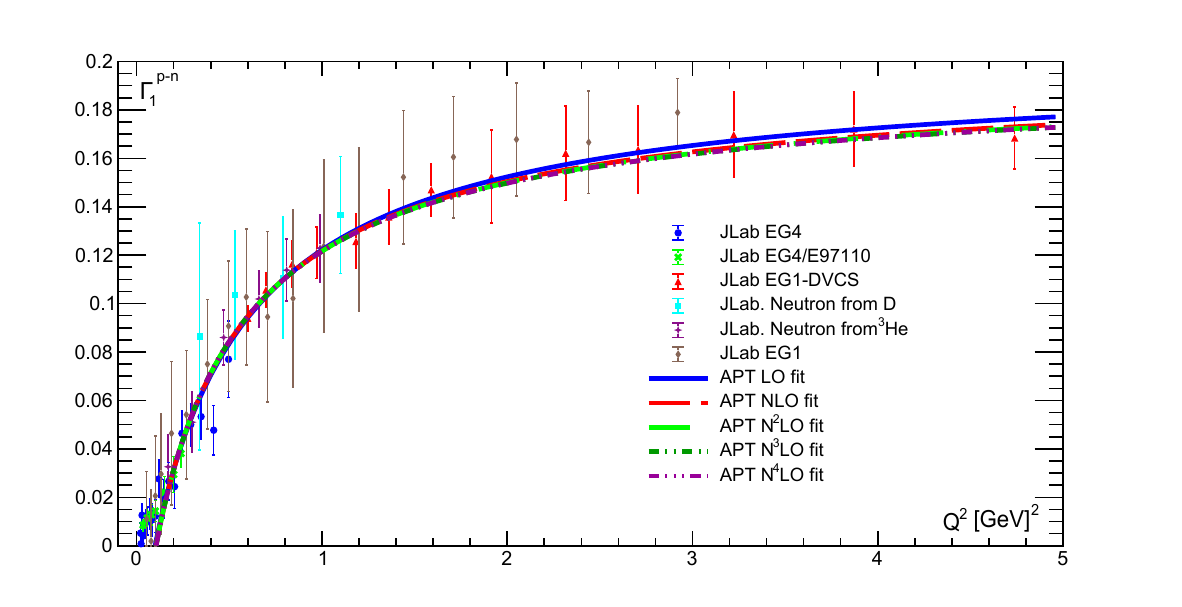}
\caption{
  \label{fig:APT}
  The results for $\Gamma_1^{p-n}(Q^2)$ in the first  four
  orders of APT.
    }
\end{figure}

The fitting results  of experimental data (see \cite{Deur:2021klh}-\cite{ResonanceSpinStructure:2008ceg})
obtained only with
statistical uncertainties 
are
shown in Figs.
\ref{fig:PT} and \ref{fig:APT}.
For the fits we use $Q^2$-independent $M^2_{\rm BS}$ and $\hat{\mu}^{\rm BS}_4$
and the two-twist part shown in Eqs. (\ref{Gpn.mOPE}) and (\ref{Gpn.MA}) 
for regular PT and APT, respectively.

As it can be seen in Fig. \ref{fig:PT}, with the exception of LO, the results obtained using conventional coupling
are very poor.
Moreover, the
discrepancy 
in this case increases with the order of PT (see also
\cite{Pasechnik:2008th,Khandramai:2011zd,Ayala:2017uzx,Ayala:2018ulm} for similar analyses).
The LO results
describe experimental points 
  relatively well, since the value of
$\Lambda_{\rm LO}$ is quite small compared to other $\Lambda_{i}$, and
disagreement with the data begins at lower values of $Q^2$ (see \cite{Gabdrakhmanov:2023rjt,Gabdrakhmanov:2024bje,Gabdrakhmanov:2025afi} and discussions therein)
below).
Thus, using  the ``massive'' twist-four form (\ref{Gpn.mOPE}) does not improve these results, since with
$Q^2 \to \Lambda_i^2$ conventional couplings
become singular, which leads to large and negative results for
the twist-two part (\ref{DBS}). So, as the PT order increases, ordinary couplings
become singular for ever larger $Q^2$ values, while Bjorken sum rule tends to negative values for ever
larger $Q^2$ values.

In contrast, our results obtained for different APT orders are practically equivalent:
the corresponding curves become indistinguishable when $Q^2$ approaches 0 and slightly different everywhere else. As can be seen in Fig. \ref{fig:APT}, the fit quality is pretty high,
which is demonstrated 
by the values of the corresponding $\chi^2/({\rm d.o.f.})$ (see Table \ref{Tab:BSR}), where the fitted values of $\hat{\mu}^{\rm BS}_{\rm{MA},4}$ and $M_{\rm BS}$
are also presented.

\subsection{Results for  Gross-Llewellyn-Smith sum rule}

By analogy with the previous subsection,
we perform a fitting of experimental data for the GLS sum rule within
the framework of standard and analytical QCD with the
``massive'' forms of the twist-four terms (see Eqs. (\ref{Gpn.mOPEa})
and (\ref{Gpn.MAa}), respectively).
\footnote{A similar analysis of the GLS sume rule can be found in \cite{Milton:1998ct}.}

\begin{figure}[!htb]
\centering
\includegraphics[width=0.98\textwidth]{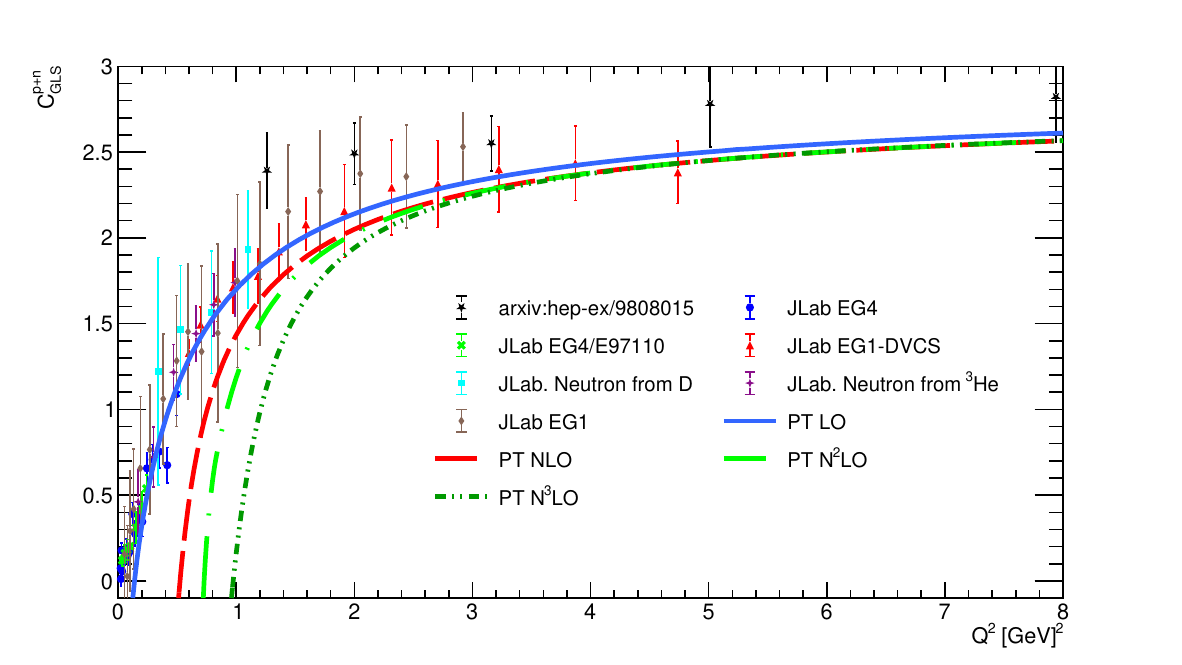}
\caption{
  \label{fig:PT2}
  As in Fig. 6 for the GLS sum rule
}
\end{figure}

As in the Bjorken sum rule case (see \cite{Gabdrakhmanov:2024bje,Gabdrakhmanov:2023rjt}) and the previous subsection),
in Figs. \ref{fig:PT2}
we see a lack of agreement between
the standard QCD predictions and the experimental data.
Moreover, we see that the discrepancy increases with increasing PT order. This is the same as in the Bjorken sum rule case, and the reason is the same.
With increasing PT order, the Landau pole of the strong coupling constant shifts toward higher $Q^2$ values. Thus, the results for GLS sum rule shifts toward negative values,
since the PT corrections have a negative sign.

\begin{table}[!htb]
\begin{center}
\begin{tabular}{|c|c|c|c|}
\hline
& $M_{\rm GLS}^2$ [Gev$^2$] for APT&  $\hat{\mu}^{\rm GLS}_{\rm{MA},4}$  for APT & $\chi^2/({\rm d.o.f.})$ for APT \\
& (for PT) & (for PT) &  (for PT)\\
\hline 
LO & 0.448 $\pm$ 0.041 & -2.99 $\pm$ 0.13 & 0.75  \\
& (0.209 $\pm$ 0.124) & (-4.87 $\pm$ 2.34) & (0.46)  \\
 \hline
  NLO & 0.377 $\pm$ 0.040 & -2.95 $\pm$ 0.16 & 0.80  \\
& (0.159 $\pm$ 2.418) & (-2.60 $\pm$ 37.08) & (0.35)  \\
 \hline
N$^2$LO & 0.358 $\pm$ 0.039 & -3.00$\pm$ 0.17 & 0.82  \\
 & (0.163 $\pm$ 8.911) & (-2.31$\pm$ 120.92) & (0.35)  \\
 \hline
 N$^3$LO & 0.354 $\pm$ 0.038 & -3.03 $\pm$ 0.17 & 0.82  \\
  & (0.116 $\pm$ 7.282) & (-2.85 $\pm$ 172.51) & (0.43)  \\
 \hline
\end{tabular}
\end{center}
\caption{%
  The values of the fit parameters in (\ref{Gpn.mOPEa}), i.e. for GLS sum rule with massive twist-4 term.
}
\label{Tab:BSR}
\end{table}

\begin{figure}[!htb]
\centering
\includegraphics[width=0.98\textwidth]{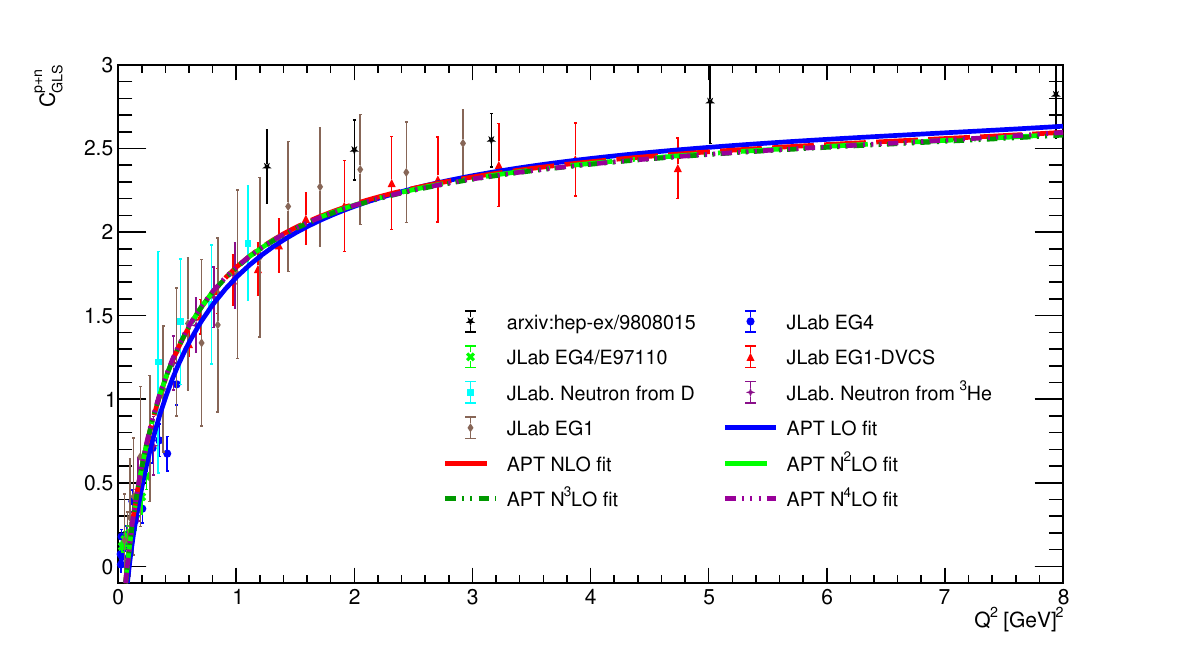}
\caption{
  \label{fig:APT1}
  As in Fig. 7 but for the GLS sum rule.
}
\end{figure}

The corresponding results for the APT case are presented in Table 2. They demostrate good $\chi^2/DOF$ value.
So, we have good agreement between the QCD predictions and the experimental data (see Fig. \ref{fig:APT1}).
 Since the number of experimental points for
 the GLS sum rule is small (see \cite{Kim:1998kia}), we have added in Figs. \ref{fig:PT2}  and \ref{fig:APT1}
 the experimental BSR data rescaled by a factor of $18/g_A$. 
We see that the experimental data for the GLS sum rule lie slightly above these rescaled experimental points for Bjorken sum rule. This observation is consistent with the results of
\cite{Londergan:2010cd}, where the possibility was discussed that the experimental data for the GLS sum rule should be smaller than those published in \cite{Kim:1998kia}
due to some unconsidered corrections.

\section{Conclusion}

We have reviewed the framework of QCD with analytic coupling, focusing on the Minimal Approach (MA) and its fractional extension (FAPT). We presented compact expressions for the MA couplings valid to any order in perturbation theory, based on the $1/L$-expansion. The key feature of this expansion is that it remains valid for all values of $Q^2$, not just the high-$Q^2$ region.

As can be clearly seen, all our results
have a compact form and do not contain complicated special functions,
  such as the Lambert $W$-function \cite{Magradze:1999um}, which already appears at the two-loop order as an exact solution to the usual coupling
  and which was used
  to evaluate  MA couplings
  in \cite{Bakulev:2012sm}.

As applications, we
considered the Bjorken and Gross-Llewellyn Smith sum rules,
$\Gamma_{1}^{p-n}(Q^2)$ and $C_{GLS}^{p+n}(Q^2)$, in the framework of MA and
perturbative
QCD and obtained results similar to those obtained in previous studies
\cite{Pasechnik:2008th,Khandramai:2011zd,Ayala:2017uzx,Ayala:2018ulm,Gabdrakhmanov:2023rjt}
for the first 4 and 3 orders of PT, respectively.
The results based on the conventional PT do not agree with the experimental data. For some $Q^2$ values, the PT results become negative, since the
high-order corrections are large and enter the twist-two term with a minus sign.
APT in the minimal version leads to a good agreement with experimental data when we used the ``massive'' version (\ref{Gpn.MA}) for
the twist-four contributions.



The results demonstrate the utility of analytic coupling methods for extending the applicability of perturbative QCD into the infrared region.

\section*{Acknowledgments}

Authors thank Konstantin Chetyrkin, Andrei Kataev and Sergey Mikhailov for careful discussions.
One of us (I.A.Z.)
is supported by Fellowship ANID Beca de Doctorado Nacional No. 21250067.


\end{document}